\documentclass[preprint,proceedings]{rmaa}

\SetYear{2006}
\SetConfTitle{Massive Stars: Fund.\ Param.\ and CS Interactions}

\title{Galactic Twins of the Ring Nebula Around SN1987A and a Possible
LBV-like Phase for Sk-69 202}

\author{Nathan Smith\altaffilmark{1}}
\altaffiltext{1}{Astronomy Dept., UC Berkeley}
\suppressfulladdresses
\listofauthors{Nathan Smith}
\indexauthor{Smith, N.}

\addkeyword{circumstellar matter}
\addkeyword{stars: evolution}
\addkeyword{stars: winds, outflows}
\addkeyword{supernovae: individual (SN1987A)}
\begin{document}
\maketitle

\boldabstract{Some core-collapse supernovae show clear signs of
  interaction with dense circumstellar material that often appears to
  be non-spherical.  Circumstellar nebulae around supernova
  progenitors provide clues to the origin of that asymmetry in
  immediate pre-supernova evolution.  Here I discuss outstanding
  questions about the formation of the ring nebula around SN1987A and
  some implications of similar ring nebulae around Galactic B
  supergiants.  Several clues hint that SN1987A's nebula may have been
  ejected in an LBV-like event, rather than through interacting winds
  in a transition from a red supergiant to a blue supergiant.}

It is commonly assumed that bipolar nebulae consist of slow ambient
material that is swept-up by the faster wind of the hot supergiant.
To create the bipolar shape, the surrounding slow wind must have an
equatorial density enhancement (i.e. a disk); the consequent mass
loading near the equator slows the expansion and gives rise to a
pinched waist and bipolar structure.  However, it's unclear how the
required pre-existing disk could have been formed. One does not
normally expect RSG or AGB stars to rotate rapidly, so a disk-shedding
scenario probably requires the tidal influence of a companion during
prior evolutionary phases in order to add sufficient angular
momentum. In the case of SN~1987A, a binary merger would be required
for this particular scenario to work (Collins et al.\ 1999).  However,
there are reasons to question this binary merger scenario for the
formation of SN1987A's nebula:

1.  A merger model followed by a transition from a RSG to BSG requires
that these two events be synchronized with the supernova event itself,
requiring that the best observed supernova in history also happens to
be a very rare event.  One could easily argue, though, that the merger
and the blue loop scenario might not have been invented if SN1987A had
occurred in a much more distant galaxy where it would not have been so
well-observed (i.e. we wouldn't know about the bipolar nebula or its
BSG progenitor).

2.  After the RSG swallowed a companion star and then contracted to
become a BSG, it should have been rotating at or near its critical
breakup velocity.  Even though pre-explosion spectra (Walborn et al.\
1989) do not have sufficient resolution to measure line profiles,
Sk--69$\arcdeg$202 showed no evidence of rapid rotation (e.g., like a
B[e] star spectrum).

3.  Particularly troublesome is that this merger and RSG/BSG
transition would need to occur twice. From an analysis of light echoes
for up to 16 yr after the supernova, Sugerman et al.\ (2005) have
identified a much larger bipolar nebula with the same axis orientation
as the more famous inner triple ring nebula.  If a merger and RSG/BSG
transition are to blame for the bipolarity in the triple-ring nebula,
then what caused it in the older one?

\begin{figure}[!t]
  \includegraphics[width=\columnwidth]{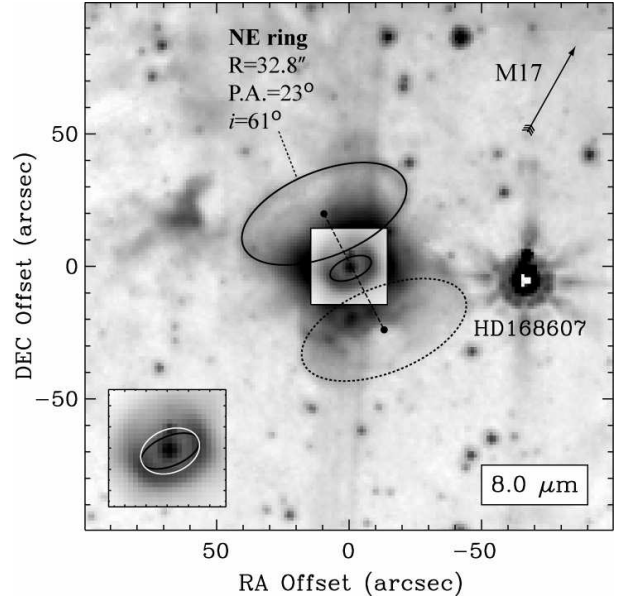}
  \caption{An 8 $\mu$m Spitzer/IRAC image of the LBV candidate
  HD168625 from Smith (2007).  It shows a nebula with a geometry very
  much like that around SN1987A, but in this case the bipolar shape
  probably originated in the ejection by the central LBV star and not
  from interacting winds.}
  \label{fig:simple}
\end{figure}

Perhaps a more natural explanation would be that Sk--69$\arcdeg$202
suffered a few episodic mass ejections analogous to LBV eruptions in
its BSG phase (see Smith 2007).  The B[e] star R4 in the Small
Magellanic Cloud may offer a precedent at the same luminosity as the
progenitor of SN1987A; R4 is consistent with a 20 M$_{\odot}$
evolutionary track, and it experienced an LBV outburst in the late
1980's (Zickgraf et al.\ 1996).  R4 also has elevated nitrogen
abundances comparable to the nebula around SN~1987A.

\begin{figure}[!t]
  \includegraphics[width=\columnwidth]{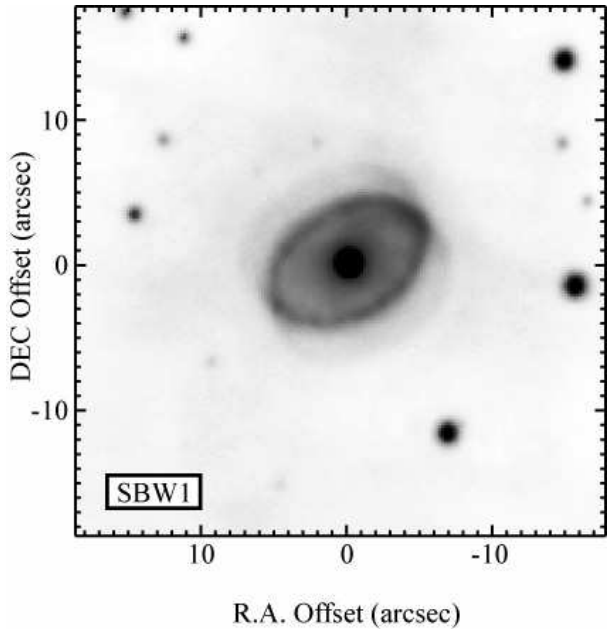}
  \caption{An H$\alpha$ image of the ring nebula SBW1 in the Carina
  Nebula, surrounding a B1.5 Iab supergiant.  It has the same radius
  as the ring around SN1987A and the star has the same luminosity as
  the progenitor of SN1987A, but it has solar N abunance, indicating
  that it has not yet been a RSG.}
  \label{fig:simple}
\end{figure}

We can gain further insight to the formation of SN1987A's ring nebula
and its pre-SN evolutionary state by studying analogs of it around
massive stars in our own Galaxy.  Three close analogs in the Milky Way
are currently known:

{\bf Sher 25 in NGC3603:} HST images of this B1.5 supergiant show a
remarkable equatorial ring with the same radius as the one around
SN1987A, plus bipolar ejecta (Brandner et al.\ 1997).  Although the
nebula has moderate N-enrichment, Smartt et al.\ (2002) find that the
N abundance is too low to be the result of post-RSG evolution.  In
fact, the stellar luminosity is above the limit where no RSGs are
seen.  Thus, the nebula around Sher 25 did not form from interacting
winds during a RSG-BSG transition.

{\bf HD168625 near M17:} This LBV candidate has a luminosity closer to
the progenitor of SN1987A than Sher 25.  Its nebula has an equatorial
ring, and it is the only object known so far to also show polar rings
like SN1987A (Fig.\ 1; see Smith 2007).  It is therefore our Galaxy's
closest analog to the progenitor of SN1987A.  Its LBV status is
interesting, since LBVs are known to have eruptive episodes of high
mass loss (e.g., Smith \& Owocki 2006) and are often surrounded by
bipolar nebulae.  Based on the properties of the nebula, I have argued
(Smith 2007) that the nebula was probaby ejected as an LBV.

{\bf SBW1 in the Carina Nebula:} This equatorial ring nebula (Fig.\ 2)
also has the same 0.2pc radius as the one around SN1987A, and the
central B1.5 supergiant has essentially the same luminosity as
Sk-69$\arcdeg$202.  It is seen in the Carina Nebula, but it is
probably more distant, at $\sim$7kpc (Smith et al.\ 2007).  Its nebula
shows no evidence for N-enrichment; the N abundance is roughly solar
(Smith et al\ 2007).  Thus, the star has never been a RSG either.

Of the three examples of ring nebulae around BSGs that are our
Galaxy's closest known analogs to the nebula around the progenitor of
SN1987A, two could not have been red supergiants because of their
chemical abundances, and one was ejected as an LBV.  Thus, of the
three examples known, {\it none} were formed by interacting winds
during a RSG to BSG transition.  This proves that there must be some
other physical mechanism that can eject equatorial rings and bipolar
nebulae.  The best candidate is an intrinsically bipolar ejection by a
rotating LBV, or an episodic mass ejection analogous to LBV outbursts.
The star does not necessarily need to have a high angular velocity, as
the effects of rotational shaping can be enhanced in a star with even
moderate angular speed if it is near the Eddington limit.  This also
hints that SN1987A and other type II SNe with circumstellar material
did not necessarily transition recently from the RSG phase; instead,
they may have been in an LBV-like phase before explosion.  If LBVs can
be SN progenitors, it puts a rather embarassing spotlight on our
current lack of an explanation for the LBV instability.

\end{document}